\documentclass[aps,twocolumn,pre,showpacs,groupedaddress,amsmath,amssymb,superscriptaddress]{revtex4}

\usepackage{bm}
\usepackage{amsmath}  
\usepackage{graphicx}   
\usepackage{verbatim}   
\usepackage{color}         
\usepackage{subfigure}  
\usepackage{hyperref}   
\usepackage{soul}
\usepackage{url}

\def \dt {\partial_{t}}
\def \dta {\partial_t^{(1)}}
\def \dtb {\partial_t^{(2)}}
\def \dtone {\partial_t^{(1)}}
\def \gradc {{\bf{\nabla}}_\mathrm{c}}
\def \grad {{\bf{\nabla}}}

\def \dr {\partial_r}

\def \dz {\partial_z}

\def \da {\partial_\alpha}
\def \db {\partial_\beta}
\def \daone {\partial_\alpha^{(1)}}
\def \dbone {\partial_\beta^{(1)}}
\def \dgone {\partial_\gamma^{(1)}}
\def \ddone {\partial_\delta^{(1)}}

\def \feq {f^{\mathrm{eq}}}
 
\def \cia {c_{i\alpha}}
\def \cib {c_{i\beta}}
\def \cig {c_{i\gamma}}
\def \cid {c_{i\delta}}
\def \cie {c_{i\eta}}
\def \cir {c_{ir}}
\def \ciz {c_{iz}}
\def \fiz {f_i^{(0)}}
\def \fia {f_i^{(1)}}
\def \fib {f_i^{(2)}}
\def \hia {h_i^{(1)}}
\def \hib {h_i^{(2)}}
\def \cs {c_s}

\def \cylinderradius {R_\mathrm{C}}
\def \systemsize {N_z \times N_r}
\def \dropradius {R_\mathrm{D}}
\def \dropcenter {(z_0, r_0)}
\def \interactionparameter {G}

\newcommand{\density}[1]{\rho_{#1}}
\newcommand{\viscosity}[1]{\nu_{#1}}

\def \st {\gamma_\mathrm{lv}}
\def \stsc {\tilde{\gamma}_\mathrm{lv}}

\def \ueq {{\textbf{u}}^{\mathrm{eq}}}

\newcommand{\REM}[1]{}
\newcommand{\COM}[1]{}
\newcommand{\ADD}[1]{}

\newcommand{\vect}[1]{\textbf{#1}}

\newcommand{\Order}[1]{\mathcal{O}{#1}}

\begin{document}

\title{Axisymmetric Multiphase Lattice Boltzmann Method} 

\author{Sudhir Srivastava} 
\affiliation{Department of Applied
  Physics, Department of Mathematics and Computer Science and
  J.M. Burgerscentrum, Eindhoven University of Technology, P.O. Box 513, 5600 MB
  Eindhoven, The Netherlands} 
\author{Prasad Perlekar} 
\affiliation{Department of Applied
  Physics, Department of Mathematics and Computer Science and
  J.M. Burgerscentrum, Eindhoven University of Technology, P.O. Box 513, 5600 MB
  Eindhoven, The Netherlands} 
\affiliation{\ADD{Centre for Interdisciplinary Sciences, TIFR, 
    21 Brundavan Colony, Narsingi, Hyderabad 500075, India} }
\author{Jan H.M. ten Thije Boonkkamp} 
\affiliation{Department of Mathematics and Computer Science and
  J.M. Burgerscentrum, Eindhoven University of Technology, P.O. Box 513, 5600 MB
  Eindhoven, The Netherlands} 
\author{Nishith Verma}
\affiliation{Department of Chemical Engineering, Indian Institute of
  Technology Kanpur, Kanpur 208016, India}
\author{Federico Toschi} 
\affiliation{Department of Applied
  Physics, Department of Mathematics and Computer Science and
  J.M. Burgerscentrum, Eindhoven University of Technology, P.O. Box 513, 5600 MB
  Eindhoven, The Netherlands} 
\affiliation{IAC, CNR, Via dei Taurini 19, I-00185 Roma, Italy }

\date{\today}

\begin{abstract}
  A novel lattice Boltzmann method (LBM) for axisymmetric multiphase
  flows is presented and validated. The novel method is capable of
  accurately modelling flows with variable density.  We develop the the
  classic Shan-Chen multiphase model [Physical Review E {\bf{47}}, 1815
    (1993)] for axisymmetric flows. The model can be used to efficiently
  simulate single and multiphase flows.  The convergence to the
  axisymmetric Navier-Stokes equations is demonstrated analytically by
  means of a Chapmann-Enskog expansion and numerically through several
  test cases. In particular, the model is benchmarked for its accuracy
  in reproducing the dynamics of the oscillations of an axially
  symmetric droplet and on the capillary breakup of a viscous liquid
  thread. Very good quantitative agreement between the numerical
  solutions and the analytical results is observed.
\end{abstract}

\pacs{47.11.-j 05.20.Dd 47.55.df 47.61.Jd}

\maketitle

\section{Introduction}
Multiphase flows occur in a large variety of phenomena, in nature and
industrial applications alike. In both type of applications it is
often necessary to accurately and efficiently simulate the
dynamics of interfaces under different flow conditions.  A
paradigmatic industrial application concerns the formation of small
ink droplets from inkjet printer nozzles \cite{Wijshoff2010}. When
both flow geometry and initial conditions display axial symmetry,
one expects that the flow will preserve that symmetry at any later
time.  Under such conditions it is advantageous to employ
numerical methods capable of exploiting the symmetry of the
problem. The computational costs of a 3-dimensional (3D) axisymmetric simulation is
very close to that of a 2-dimensional (2D), presenting thus a considerable
advantage over fully 3D simulations. 
When one deals with multiphase methods characterized by
diffused interfaces, such as the ones common in the lattice Boltzmann method, 
the availability of additional computational resources allows one to decrease the
interface width with respect to the other characteristic length-scales in the 
problem. The possibility to get closer to the ``sharp-interface''
limit has thus a direct impact on the accuracy of the numerical
solutions for diffuse interface multiphase solvers.

The lattice Boltzmann method (LBM) \cite{Succi_book2001} has been widely
employed to study multiphase flows in complex geometries under both
laminar and turbulent flow conditions \cite{Chen1998}. In recent years
several implementations of axisymmetric LBM for single-phase systems
have been proposed
\cite{Halliday2001,Reis2007,Huang2009,Guo2009,Li2010,Chen2008}, while, in comparison,
relatively little attention has been devoted to the case of the
multiphase flow \cite{Premnath2005,Mukherjee2007}.

The aim of the present paper is to introduce a novel, accurate and
efficient algorithm to study generic axisymmetric, density-varying
flows and in particular multiphase flows. The proposed algorithm is
easy to implement, is accurate and its multiphase model builds upon
the widely used Shan-Chen model \cite{Shan1993,Shan1994}. One
particular advantage of having the axisymmetric implementation of
the Shan-Chen model is that it allows one to retain the same parameters of
the fully 3D model (e.g., coupling strength, surface tension and phase
diagram) thus allowing to easily switch between axisymmetric and
full 3D Shan-Chen investigations, according to what is needed.

The manuscript is organized as follows. In Section~\ref{sec:MODEL} we
present the new lattice Boltzmann method. In
Section~\ref{sec:validation-sp} and Section~\ref{sec:validation-mp} we present the 
results of several benchmarks of the method against single and multiphase flows,
respectively. In Section~\ref{sec:conclusion} conclusions are
drawn. The derivation of the additional terms for the axisymmetric LBM model 
is presented in Appendix \ref{sec:appA}. 

\section{MODEL}\label{sec:MODEL}
\subsection{Multiphase lattice Boltzmann method}
In this section we introduce the notation and quickly recall the
basics of the Shan-Chen LBM; in particular we
focus on the 2D and nine velocities (D2Q9) Shan-Chen (SC)
model for multiphase flow \cite{Shan1993,Shan1994}. 
The LBM is
defined on a Cartesian, 2D lattice together with the
nine velocities, $\vect{c}_i$, and distribution functions, $f_i$. The time
evolution of the populations is a combination of free streaming and
collisions:
\begin{equation} \label{eq:lbm2d}
  f_i(\vect{x}+ \vect{c}_i\delta t, t+\delta t) = f_i(\vect{x},t) -
  \frac{1}{\tau} \big( f_i(\vect{x},t)-\feq_i(\rho,\ueq)
  \big).
\end{equation}
In the particular case of Eq.~(\ref{eq:lbm2d}), we have further made
use of the so-called BGK approximation where a single relaxation time,
$\tau$, is used to relax the population distributions towards the
equilibrium distributions, $\feq_i$. In our notations the relaxation parameter, $\tau$, is
scaled by the time step, $\delta t$. The kinematic viscosity of the
fluid, $\viscosity{}$, is related to the relaxation parameter, $\tau$,
by ${\viscosity{} = c_s^2 \delta t \left(\tau -
  0.5\right)}$, where $c_s = \sqrt{1/3}$ is the speed of sound for the D2Q9 model. The fluid density is defined as
$\rho=\sum_{i}f_i$.  In the SC model  the
internal/external force, $\vect{F}$, is added to the system by
shifting the equilibrium velocity as \cite{Shan1993,Shan1994}:
\begin{equation}\label{eq:ueq}
  \ueq =\frac{1}{\rho}\left(\sum_i {\bm c}_if_i + \tau\,\delta t\,\vect{F}\right),
\end{equation} 
while the hydrodynamic velocity is defined as 
\begin{equation}\label{eq:uhydro}
  \vect{u}=\frac{1}{\rho} \left(\sum_i {\bm c}_if_i +\frac{\delta t}{2} \vect{F}\right).
\end{equation}
The short-range (first
neighbors) Shan-Chen force, $\vect{F}({\bf x})$, at position ${\bf x}$ is defined as
\begin{equation}\label{eq:scforce}
  \vect{F}({\bf x})  = -G \psi({\bf x}) \sum_i  W_i  \psi({\vect{x} + \vect{c}_i \delta t})\vect{c}_i,
\end{equation}
where $G$ is the interaction strength, and the $W_i$'s are the lattice
dependent weights. The  density functional is $\psi\big(\rho(\vect{x})\big) =
\rho_0\big(1 - \exp(-\rho(\vect{x}) /\rho_0)\big)$ where $\rho_0$ is a reference
density and is equal to unity for the results presented in this manuscript. 
From this setting it follows that the bulk pressure, ${p_{\mathrm{NI}}}$,
and pressure tensor, $P_{\alpha\beta}$ $(\mathrm{for\,} \delta t = 1)$ are given by:
\begin{equation}\label{eq:pressure}
  p_{\mathrm{NI}}  = c_s^2 \rho +  \frac{c_s^2G}{2}\psi(\rho)^2,
\end{equation}
\begin{eqnarray}\label{eq:pressure-tensor}
  P_{\alpha\beta}  &=& \bigg(c_s^2\rho+ \frac{c_s^2G}{2}\psi^2 + \frac{c_s^4G}{2}\psi \gradc^2 \psi+ \frac{c_s^4G}{4}|\gradc \psi |^2\bigg)\delta_{\alpha\beta} \nonumber\\
  &&- \frac{c_s^4G}{2}\da \psi\db \psi + \bigg(\tau - \frac{1}{2}\bigg)^2 \frac{1}{\rho} F_\alpha F_\beta,
\end{eqnarray}
respectively, and the surface tension $\st$ is given by 
\begin{equation}\label{eq:st}
  \st = -\frac{Gc_s^4}{2}\int_{-\infty}^{\infty} (\gradc \psi \cdot \hat{\vect{n}})^2 \,dn,
\end{equation}
where $\delta_{\alpha\beta}$ is the Kronecker delta
function, $\hat{\vect{n}}$ is the unit vector normal to the interface and $\gradc$ and $\gradc^2$ are the 2D Cartesian gradient and Laplacian operator, respectively
(see \cite{He2002, Benzi2006,Shan1994} for details). Varying the interaction strength, $G$, and 
choosing an average density, it can be shown that the system can
phase-separate and model the coexistence of a liquid and its
vapor. This multiphase system is characterized by a larger density in
the liquid phase and a lower density in the vapor phase and by a
surface tension at the interface separating the two phases. For the scheme proposed in \cite{Shan1993,Shan1994} the surface tension given by Eq.~(\ref{eq:st})
should have a $\tau-$correction term, which is due to the last term of Eq.~(\ref{eq:pressure-tensor}) and hence the surface tension is given by 
\begin{equation}\label{eq:stc}
  \stsc = -\frac{Gc_s^4}{2}\int_{-\infty}^{\infty} (\gradc\psi \cdot \hat{\vect{n}})^2 \,dn + \bigg(\tau - \frac{1}{2}\bigg)^2 \int_{-\infty}^{\infty} (\vect{F}\cdot \hat{\vect{n}})^2\frac{1}{\rho} \,dn.
\end{equation}
The $\tau-$correction term in Eq.~(\ref{eq:stc}) is the consequence of the choice of the scheme used for adding the external/internal forces in LBE, for example, if we use the force incorporation scheme proposed in \cite{Guo2002} the surface tension should not have the $\tau-$correction.

\subsection{Axisymmetric Navier-Stokes equations}
When the boundary conditions, the initial configuration and all
external forces are axisymmetric, one does expect that the
solution of the Navier-Stokes (NS) equations will preserve the axial symmetry at any
later time. 
The continuity and NS equations in the cylindrical coordinates $(z,r,\theta)$, in absence of external forces 
reads:
\begin{align}\label{eq:axis_cty}
  \dt \rho + \db(\rho u_{\beta}) & = -r^{-1}\rho u_r, 
\end{align}
and
\begin{subequations}\label{eq:axis_ns}
  \begin{eqnarray}
    \rho (\dt u_z + u_{\beta} \db u_z ) & =& -\dz p + \db\big(\mu(\db u_z + \dz u_{\beta})\big) \nonumber\\
    &&+ r^{-1}\mu (\dr u_z + \dz u_r), \\
    \rho (\dt u_r + u_{\beta} \db u_r ) & =& -\dr p + \db\big(\mu(\db u_r + \dr u_{\beta})\big)  \nonumber\\
    &&+ 2\mu\dr\left(r^{-1}u_r\right), 
  \end{eqnarray}
\end{subequations}
respectively, where ${\mu=\nu\rho}$, is the dynamic viscosity and $\nu$ is the kinematic viscosity of the fluid. The index $\beta$ runs over the set $\{z,r\}$, and when an index 
appears twice in a single term it represents the standard Einstein summation convention. In principle an axisymmetric flow may have an azimuthal component of the velocity field, $u_{\theta}$. In Eqs.~(\ref{eq:axis_cty}) and (\ref{eq:axis_ns}) we assume that the flows that we consider
have no swirl $,i.e.,{u_\theta = 0}$, and that other hydrodynamic variables are independent of $\theta$. We can thus write, $u_r = u_r(z,r;t)$, $u_{\theta}=0$, $u_z=
u_z(z,r;t)$ and $\rho = \rho(z,r;t)$. 

The axisymmetric version of the continuity and NS equations have been
recast in a form, Eqs.~(\ref{eq:axis_cty}) and (\ref{eq:axis_ns}), to easily highlight the similarities with respect to
2D flows in a $(z, r)$-plane.

Our approach employs a 2D LBM to solve for the two-dimensional part of
the equations and explicitly treat the additional terms.

The continuity equation differs from the purely
2D because of the presence of a source/sink term on the right hand side of Eq.~(\ref{eq:axis_cty}); this term is responsible for a locally increasing
mass whenever fluid is moving towards the axis, and for
decreasing mass, when moving away. The physical role of this term is to
maintain 3D mass conservation (a density $\rho$ at a distance $r$ must
be weighted with a $2\pi r$ factor).

The NS equations have also been rewritten in a way to highlight the 2D
equations. The additional contributions that make the 3D axisymmetric
equations differ from the 2D ones are the terms ${r^{-1}\mu (\dr u_z + \dz u_r)}$ and ${ 2\mu\dr (r^{-1}u_r)}$ 
on the right hand side of the Eqs.
(\ref{eq:axis_ns}). In our LBM model these terms are also explicitly
evaluated and added as additional forcing terms.

The idea to model the 3D axisymmetric LBM with a 2D LBM supplemented
with appropriate source-terms has already been employed in a number of
studies, for single-phase axisymmetric LBM models
\cite{Halliday2001,Reis2007,Reis2007a,Reis2008} and for multiphase LBM
as well \cite{Premnath2005, Mukherjee2007}.
Here we will develop an axisymmetric version of the Shan-Chen model
\cite{Shan1993,Shan1994}. 

From here onwards we will use the following notations: ${\vect{x} = (z, r)}$, ${\vect{u} = (u_z, u_r)}$
and ${\gradc = (\partial_z, \partial_r)}$, where $z$-axis is the horizontal axis and $r$-axis is the vertical axis.

\subsection{LBM for axisymmetric flow}
\begin{figure}[!ht]
  \includegraphics[width=0.4\textwidth]{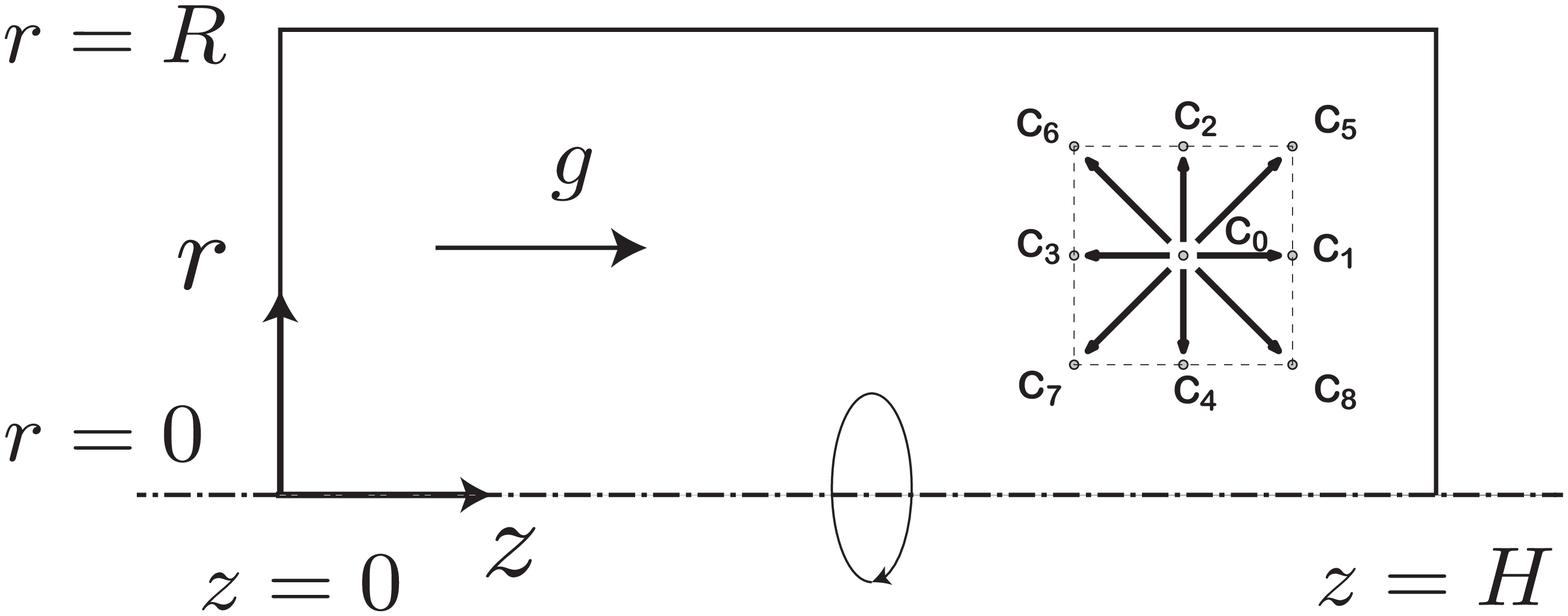}
  \caption{Schematics of the axisymmetric geometry, eventually in presence of gravity or an external force aligned with the z-axis. Schematics also shows the lattice velocities for the D2Q9 model.}
  \label{fig:sch-HPFlowtest}
\end{figure}
The first step in deriving a LBM for axisymmetric multiphase
flows is to derive a model that can properly deal with density
variations. In particular, the LBM should recover the
axisymmetric continuity Eq.~(\ref{eq:axis_cty}) and NS
Eqs.~(\ref{eq:axis_ns}) by means of a Chapman-Enskog (CE) expansion
in the long-wavelength and long-timescale limit. In order to
derive such a model we start from the 2D LBM with the addition of
appropriate space- and time-varying microscopic sources $h_i$ (see
also \cite{Halliday2001,Reis2007,Reis2007a,Reis2008}).  We employ the following lattice Boltzmann
equation:
\begin{equation}\label{eq:lbm2d_mod_new}
  \begin{split}
    f_i({\vect{x}}+ {\vect{c}_i}\delta t, t + \delta t) &- f_i({\vect{x}}, t) = -\frac{1}{\tau} \big(f_i({\vect{x}},t)-\feq_i(\rho, \ueq)\big) \\
    &~~~~+ \delta t ~h_i(\vect{x} + {\vect{c}_i} \delta t /2, t + \delta t/2),
  \end{split}
\end{equation}
where the source terms $h_i$, are evaluated at fractional time
steps. It can be shown, see Appendix \ref{sec:appA}, that when the additional term
$h_i$ in Eq.~(\ref{eq:lbm2d_mod_new}) has the following form:
\begin{equation}\label{eq:condition_h}
  h_i = W_i\Big(-\frac{\rho u_r}{r} + \frac{1}{c^2_s}\big(c_{iz}H_z + c_{ir}H_r\big)\Big),
\end{equation}
with
\begin{subequations}\label{eq:H_terms}
  \begin{align}
    H_z  & = \frac{\ciz}{r}\bigg(\mu \big(\dr u_z + \dz u_r\big)  - \rho u_r u_z\bigg),    \\
    H_r & = \frac{\cir}{r}\bigg(2\mu\Big(\dr u_r - \frac{u_r}{r}\Big) -\rho u_r^2\bigg),    
  \end{align}
\end{subequations}
the CE expansion of Eq.~(\ref{eq:lbm2d_mod_new}) provides the
axisymmetric version of the continuity and of the NS Eqs.~
(\ref{eq:axis_cty}) and (\ref{eq:axis_ns}), respectively. Details on
the CE expansion are reported in Appendix \ref{sec:appA}. 
The equations introduced here are enough to describe a fluid with
variable density in axisymmetric geometry. We performed validations of
the numerical model (not reported) by observing the behavior of the
volume for the case of a droplets approaching the axis. While the 2D
volume in the system was not conserved, the properly defined 3D volume was
conserved with good accuracy.

\subsection{LBM for axisymmetric multiphase flow}
With a lattice Boltzmann method capable of handling density variations
the additional steps towards the definition of the axisymmetric
version of the SC multiphase model only consists in the correct
definition of the SC force.  The expression for the SC force in 3D is:
\begin{equation}
  \vect{F}({\bf x}) = -G \psi({\bf x})\sum_i  W_i  \psi({\vect{x} + \vect{c}_i \delta t})\vect{c}_i.
  \label{eq:sc2d}
\end{equation}
To find the lattice expression for the axisymmetric case we proceed by
passing to the continuum limit, by expressing the continuum force in
cylindrical coordinates and then by separating the 2D SC force from
the additional axisymmetric contributions.

By means of a Taylor expansion for $\psi({\bf x + c}_i\delta t)$ one
easily obtains the following continuum expression for the SC force
\cite{Benzi2006}:
\begin{eqnarray} \label{eq:sc3dc}
  \vect{F}({\bf x}) &=& -G c_s^2 \,\delta t\,\psi({\bf x}) \grad \psi({\bf x}) \nonumber\\
  &&- \frac{G}{2} c_s^4 (\delta t)^3\psi({\bf x}) \grad\big(\grad^2 \psi({\bf x})\big) \nonumber\\
  &&+ \Order((\delta t)^5).
\end{eqnarray}
The above force expression is lattice independent and holds true for any 3D
coordinate system. We restrict Eq.~(\ref{eq:sc3dc}) to the case of
axisymmetric flows by expressing both the gradient, $\grad$ and the Laplace, $\grad^2$
operators in cylindrical coordinates given by $ {\grad \equiv ( \partial_z,
  \partial_r) = \gradc}$ and ${ \grad^2 \equiv (\partial_{zz} + \partial_{rr} +
  r^{-1} \partial_r ) = \gradc^2 +  r^{-1} \partial_r}$.  Thus, in the axisymmetric case, Eq.~(\ref{eq:sc3dc}) reduces to:

\begin{eqnarray} \label{eq:sc2axsym}
  \vect{F}({\bf x}) &=& -G c_s^2 \,\delta t\,\psi({\bf x}) \gradc \psi({\bf x}) \nonumber\\
  &&- \frac{G}{2} c_s^4 (\delta t)^3\psi({\bf x}) \gradc\big(\gradc^2 \psi({\bf x})\big) \nonumber\\
  &&+ \vect{F}^{\mathrm{\gamma,sym}}({\bf x}) + \Order((\delta t)^5).
\end{eqnarray}
where
\begin{equation}\label{eq:scaxis22}
  \vect{F}^{\mathrm{\gamma,sym}}({\bf x}) = - \frac{G}{2} c_s^4 (\delta t)^3\psi({\bf x}) \gradc \big(r^{-1}\partial_r \psi({\bf x}) \big).
\end{equation}
From Eq.~(\ref{eq:sc2axsym}) we immediately recognize that the first
two terms on the right hand side are the ones that one obtains from
the Shan-Chen model in 2D. The last term in Eq.~(\ref{eq:sc2axsym}),
$\vect{F}^{\mathrm{\gamma,sym}}$, is the additional term responsible for the
three-dimensionality. This extra contributions needs to be accurately
taken into account in order to model the axisymmetric Shan-Chen
multiphase systems in 3D. In particular, this term is extremely important in order to
correctly implement a 3D surface tension force which responds to curvatures,
both along the axis and in the azimuthal direction. The two
components of the additional term can be rewritten as:
\begin{subequations}
  \begin{eqnarray}\label{eq:scAxisExtra}
    {F_z^{\mathrm{\gamma,sym}}} & = & - \frac{G}{2}c_s^4  (\delta t)^3\psi r^{-1}\partial_{zr} \psi,\\ 
    {F_r^{\mathrm{\gamma,sym}}} & = & - \frac{G}{2} c_s^4\psi  (\delta t)^3\Big(r^{-1} \partial_{rr}\psi - r^{-2}\partial_r\psi \Big). 
  \end{eqnarray}
\end{subequations}
The evaluation of the terms $F_z^{\mathrm{\gamma,sym}}$ and $F_r^{\mathrm{\gamma,sym}}$ requires an
approximation for the derivatives accurate up to order $(\delta t)^4$
or higher. Such an accuracy ensures the isotropy of the
``reconstructed'' 3D axisymmetric Shan-Chen force and thus the isotropy of the resulting surface
tension along the interface.

In our implementation we used the following isotropic 5{\it th}-order
accurate finite difference approximations on D2Q9 lattice (see FIG.~\ref{fig:sch-HPFlowtest}). For a scalar valued function
$\phi(\vect{x})$ it reads:
\begin{subequations}
  \begin{eqnarray}
    \partial_r \phi(\vect{x}) &=& \frac{1}{36}\sum_{i = 1}^8 {\Big(8\phi(\vect{x} +
      \vect{c}_i\delta t) - \phi(\vect{x} + 2\vect{c}_i\delta
      t)\Big)}c_{ir} \nonumber\\ &&+ \Order((\delta
    t)^5),\label{eq:forthorder_firstderivativex}\\ 
    \partial_{rr}  \phi(\vect{x}) &=& \frac{1}{36}\sum_{i = 1}^8{\Big(8\partial_r \phi(\vect{x} +
      \vect{c}_i\delta t) - \partial_r \phi(\vect{x} +
      2\vect{c}_i\delta t)\Big)}c_{ir} \nonumber\\ &&+
    \Order((\delta
    t)^5),\label{eq:forthorder_secondderivativex}\\ 
    \partial_{zr}  \phi(\vect{x}) &=& \frac{1}{12}\Big(- \partial_r\phi(\vect{x} + 2\vect{c}_1\delta t) + 8\partial_r\phi(\vect{x} + \vect{c}_1\delta t) \nonumber \\
    && -8\partial_r\phi(\vect{x} + \vect{c}_3\delta t) + \partial_r\phi(\vect{x} + 2\vect{c}_3\delta t)\Big) \nonumber\\
    &&+ \Order((\delta  t)^6)
    \label{eq:forthorder_mixedderivative},
  \end{eqnarray}
  where $\partial_r \phi (\vect{x}) $ in Eq.~(\ref{eq:forthorder_mixedderivative}) is approximated as
  \begin{eqnarray}
    \partial_{r}  \phi(\vect{x}) &=& \frac{1}{12}\Big(- \phi(\vect{x} + 2\vect{c}_2\delta t) + 8\phi(\vect{x} + \vect{c}_2\delta t) \nonumber \\
    && -8\phi(\vect{x} + \vect{c}_4\delta t) + \phi(\vect{x} + 2\vect{c}_4\delta t)\Big)\nonumber\\
    &&+ \Order((\delta  t)^6).
  \end{eqnarray}
\end{subequations}
\ADD{From the SC model the present axisymmtric
 implementation does inherit all advantages as well as the
 limitations. One of the limitation is the relatively small
 density contrast that can be achieved. Other multiphase LBM model,
 for which similar axisymmetric extension could similarly be worked
 out, may allow to achieve larger density contrasts. In the SC model
 both the density ratio, $\rho_l/\rho_v$, and the surface tension,
 $\gamma$, depend upon a single parameter, $G$. Decreasing the value
 of $G$ increases both the surface tension and the density ratio
 between two phases. Higher surface tension gives smaller interface
 width that leads to higher truncation error in the gradient
 approximation at the interface. This makes the numerical scheme
 unstable for too high values of $G$. As a rule of thumb, a density
 ratio of $\rho_l/\rho_v \leq 35, (G \geq -6.0)$ still ensures the
 stability of the SC model. Therefore, the current axisymmetric SC
 model, as much as the standard SC model, is limited to density ratio
 around 35.}
\subsection{Boundary conditions}
In axisymmetric flows the boundary conditions for the distribution
functions, $f_i$, need to be prescribed at all boundaries including the
axis. In our approach we impose boundary conditions before the
streaming step (pre-streaming).  We use mid-grid point specular
reflection boundary conditions on the axis \cite{sukop_book2006}, this
choice allows us to avoid the singularity due to the force terms
containing $1/r$. Mid-grid bounce-back or mid-grid specular reflection
boundary conditions are used to impose either hydrodynamic no-slip
or free-slip conditions at the other walls, respectively
\cite{sukop_book2006}.  In order to impose a prescribed velocity or
pressure at inlet and outlet boundaries, we impose the equilibrium
distribution functions, $\feq_i$, evaluated using the desired
hydrodynamic velocity and density values. For our LBM simulations we use unit time step $({\delta t = 1})$
and unit grid spacing ${(\delta z = \delta r = 1)}$, hence the length can be measured in
terms of the number of nodes. We are using symmetry boundary condition is used for the derivative evaluation in (\ref{eq:H_terms}) and (\ref{eq:scaxis22}) at the axis. For other three boundaries we impose the derivatives terms to be zero.
\section{Numerical validation for single-phase axisymmetric LBM}
\label{sec:validation-sp}

Here we present the validation of the axisymmetric LBM for
single-phase flow simulations by comparing it with analytical
solutions for the test cases: the axial flow through a tube and the outward radial flow between two
parallel discs. These two tests complement each other because they
correspond to flows parallel and orthogonal to the axis, respectively.  Both flow
problems have analytical steady state solutions that help us to
validate the accuracy of the axial and radial component of the velocity. 
All physical quantities in this manuscript, unless otherwise
stated, are reported in lattice units (l.u.), the relaxation time has
been keep fixed for all the simulations, $\tau$ = 1, and the
simulations have been carried out on a rectangular domain of size
$H \times R = \systemsize$. The steady state in the following single-phase simulations
is defined when the total kinetic energy of the system, 
${ E_{ke} = \pi \sum_{N_z} (\sum_{N_r }r \rho |\vect{u}|^2})$, becomes constant up to the machine precision.

\subsubsection{Flow through a pipe}
In this test we consider the constant-density flow of a fluid with
density, $\density{}$, kinematic viscosity, $\viscosity{}$, flowing
inside a circular pipe of radius $R$. The flow is driven by a constant
body force, ${\rho g}$, along to the axis of the pipe. The schematic
illustration of the flow geometry is presented in
FIG.~\ref{fig:sch-HPFlowtest}.
Assuming $u_r(z,r) = 0$ and no-slip condition on the inner surface $(r = R)$ of the pipe, the
steady state solution for the axisymmetric NS Eq.~(\ref{eq:axis_ns})
for this problem is given by \cite{Middleman_book1995}:
\begin{align}\label{eq:hag_pois_exsol}
  u_z(z, r)  &=     U_1\left[1-\left(\frac{r}{R}\right)^2\right],
\end{align}
where ${U_1 = u_r(z,0) = gR^2/(4\viscosity{})}$, is the maximum velocity in the pipe.
\begin{figure}[!ht]
  \centering
  \includegraphics[width=0.4\textwidth]{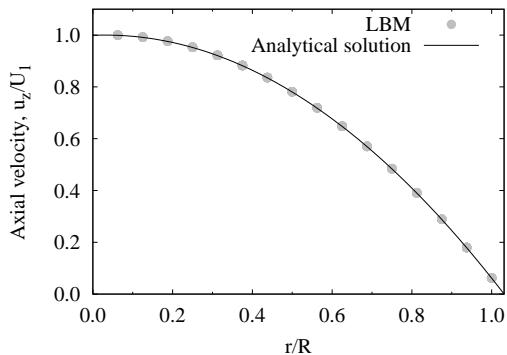}
  \caption{A comparison of the axial velocity profile as obtained form
    the LBM simulations (circles) versus the analytical solution
    (solid line) Eq.~(\ref{eq:hag_pois_exsol}). Simulation parameters:
    $\systemsize = 16 \times 16$, $R = N_r$,
    $\density{}$ = 1, $\viscosity{}$ = 0.167, $g= 10^{-5}$, $U_1 = u_z(0) = 3.84\times10^{-3}$ .}
  \label{fig:R2_HpTest}
\end{figure}

For the LBM simulation we used
the no-slip boundary condition at the inner surface of the pipe, and
periodic boundary conditions at the open ends of the pipe. The body
force $g = 10^{-5}$ is applied at each node of the simulation
domain. The LBM simulations are carried out till the simulation reaches its steady state. The result of the LBM simulation shown in
FIG.~\ref{fig:R2_HpTest} is in very good agreement with the
analytical solution in Eq.~(\ref{eq:hag_pois_exsol}). This validates the
single phase axisymmetric LBM for the case where there is no velocity
in the radial direction.

\subsubsection{Outward radial flow between two parallel discs} 
\begin{figure}[!ht]
  \includegraphics[width=0.45\textwidth]{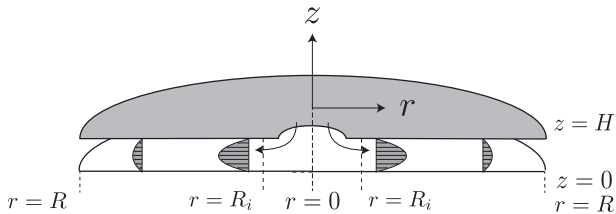}\\
  \caption{Schematics of the physical setup to study outward radial
    flow between two parallel discs. The arrows show the direction of
    the inlet mass flow. The LBM is used to simulate the flow domain
    $0\leq z \leq H$ , $R_i \leq r \leq R$.  We assume that the flow is
    fully developed for $r \geq R_i$ and hence the axial velocity $u_z$
    vanishes in this region.}
  \label{fig:sch-radialFlowtest}
\end{figure}

\begin{figure}[!ht]
  \centering
  \includegraphics[width=0.46\textwidth]{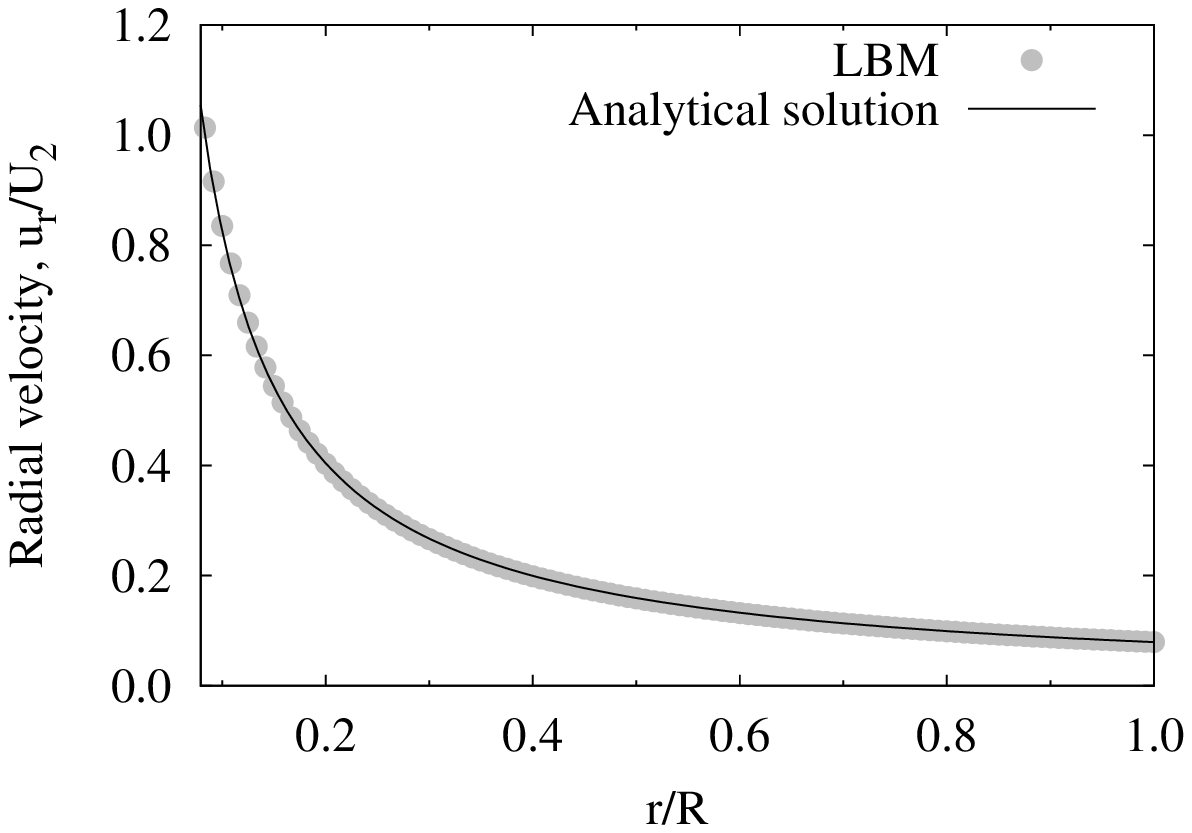}
  \includegraphics[width=0.46\textwidth]{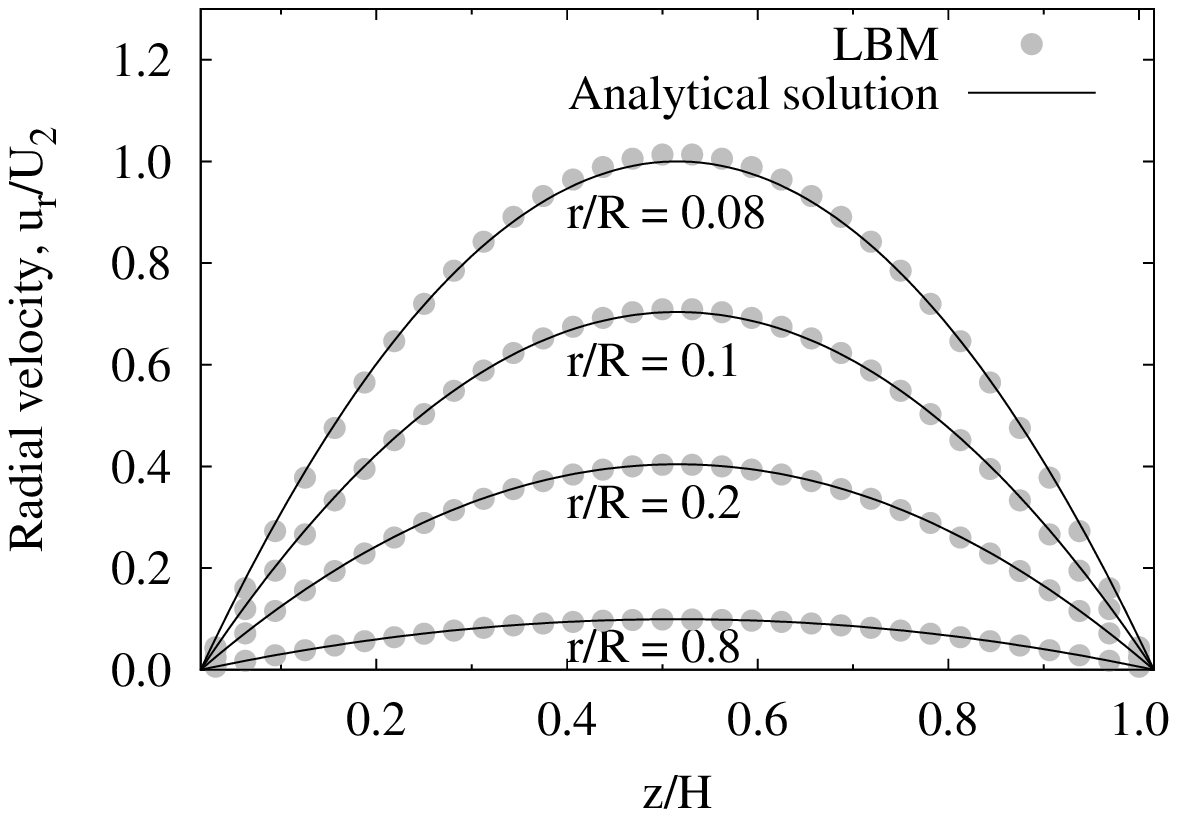}   
  \caption{A comparison of the radial velocity profile, $u_r$, as
    obtained form LBM simulations (circles) versus the analytical
    solution (solid lines) of Eq.~(\ref{eq:sol-vel-test2}).
    Simulation parameters: $\systemsize = 32 \times 120$,
    $R = N_r-0.5$, $H = N_z$, $R_i = 9.5$, $\density{}
    = 1$, $\viscosity{} = 0.167$, $Q = 0.5$ and $U_2 = 3.93 \times 10^{-4}$. Top figure shows the
    comparison at $z = 15.5$. Bottom figure from top to bottom shows the curves correspond
    to the radial distances $r/R = 0.8, 0.2, 0.1$ and  $0.08$. }
  \label{fig:R11-test1}
\end{figure}

Another important test to validate the single-phase axisymmetric LBM
is the simulation of the outward radial flow between two parallel
discs separated by a distance $H$. The schematic of the flow setup for
this problem is reported in FIG.~\ref{fig:sch-radialFlowtest}.
Assuming $u_z(z, r) = 0$ for $R_i \leq r \leq R$, the no-slip boundary condition on the discs and a constant mass
flow rate $Q$ along the radial direction, the solution of the NS
Eq.~(\ref{eq:axis_ns}) corresponding to this problem is given by \cite{Middleman_book1995}:
\begin{equation}\label{eq:sol-vel-test2}
  u_r(z,r) = -U_2\left(\frac{4R_i}{H^2}\right)\frac{z(z - H)}{r},
\end{equation}
where ${U_2 = u_r(H/2,R_i) = 3Q/(4\pi R_i H)}$. The LBM results shown in
FIG.~\ref{fig:R11-test1} are carried out for the flow domain $R_i \leq r
\leq R, 0\leq z \leq H$ and using the no-slip boundary condition along the
discs. The velocity profile given by Eq.~(\ref{eq:sol-vel-test2}) is
applied at the inlet boundary while the outlet is considered as an
open boundary.  The LBM results shown in FIG.~\ref{fig:R11-test1}
are in a very good agreement with the analytical solution
Eq.~(\ref{eq:sol-vel-test2}). This validates the single phase
axisymmetric LBM for the case of a radial velocity.

\section{Numerical validation for axisymmetric multiphase model}
\label{sec:validation-mp}
In this section we present the validation for our axisymmetric multiphase LBM for
three standard test cases: Laplace law, oscillation of a viscous drop 
and the Rayleigh-Plateau (RP) instability.

\subsubsection{Laplace test}
In this validation we compare the in-out pressures
differences for different droplet radii. 
According to the Laplace law the in-out pressure difference, 
$\Delta p$, for a droplet of radius $\dropradius$ is given by 
\begin{equation}\label{eq:laplacelaw}
  \Delta p = \frac{2\st}{\dropradius},
\end{equation}
where $\st$ is the liquid-vapor interfacial tension.  
For this validation we first estimate the value of the surface tension using 
Eq.~(\ref{eq:st}) (Guo scheme \cite{Guo2002}) and Eq.~(\ref{eq:stc}) 
(SC scheme \cite{Shan1994}) for both 2D and axisymmetric LBM. The data 
obtained from these simulations are reported in TABLE \ref{tab:sc-param}.  
\begin{table}[h!]
  \begin{tabular}{|c|c|c|c|c|}
    \hline
    \multicolumn{1}{|c|}{} & \multicolumn{2}{|c|}{SC} & \multicolumn{2}{|c|}{Guo} \\ \hline
    \interactionparameter & $\stsc^\mathrm{2D}$ & $\stsc^\mathrm{axis}$&$\st^\mathrm{2D}$ & $\st^\mathrm{axis}$\\ 
    \hline
    -4.5	& 	0.0220 & 0.0220	& 0.0135	&0.0136\\
    -5.0	&      0.0579 & 0.0579	& 0.0376	&0.0378\\
    -5.5	& 	0.0995 & 0.0996	& 0.0681	&0.0683\\
    \hline
  \end{tabular}
  \caption{Surface tension evaluated using Eq.~(\ref{eq:stc}) (column 2,3) and Eq.~(\ref{eq:st}) (column 4,5). Here $\st^\mathrm{2D}, \st^\mathrm{axis}$ denote the surface tensions obtained from 2D and axisymmetric LBM, respectively. Simulation parameters: ${ \systemsize = 1 \times 64}$, $\tau = 1$, initial interface position, $r = 32$.}
  \label{tab:sc-param}
\end{table}
Both the Guo and SC scheme are consistent with the fact that for the SC model the surface 
tension should only depends on the value of the interaction parameter, $G$. 

In the next step we do a series of axisymmetric LBM simulation for different droplet radii and measure the in-out pressure difference. 
When comparing the in-out pressure difference for a drop (Laplace test) and the pressure drop given by 
Eq.~(\ref{eq:laplacelaw}), we find that the maximum relative error in pressure difference for 
Guo scheme \cite{Guo2002} and SC scheme \cite{Shan1994} is 2\% and 20\%, respectively. This difference might 
be due to following reason.
\subsubsection{Oscillating droplet}
Here we consider the dynamics of the oscillation of an axisymmetric droplet 
in order to validate the axisymmetric multiphase LBM. We compare the
frequency of the oscillation of the droplet obtained from the LBM
simulation with the analytical solution reported in Miller and Scriven
\cite{Miller1968}. The frequency of the second mode for the oscillation
of a liquid droplet immersed in another fluid is given by:
\begin{equation}\label{eq:freq-drop-miller}
  \omega_2 = \omega_2^\ast -0.5\alpha(\omega_2^\ast)^{1/2} + 0.25\alpha^2,
\end{equation}
where 
\begin{equation*}
  \omega_2^\ast = \sqrt{\frac{24\st}{\dropradius^3 (2\density{v} + 3 \density{l})}},
\end{equation*}
and $\dropradius$ is the radius of the drop at equilibrium,
$\st$ is the surface tension, $\density{l},
\density{v}$ are the densities of the liquid and vapor phases,
respectively. The parameter $\alpha$ is given by:
\begin{equation*}
  \alpha = \frac{25 \sqrt{\viscosity{l}\viscosity{v}}\density{l} \density{v}}{\sqrt{2}\dropradius(2\density{v} + 3\density{l})(\sqrt{\viscosity{l}}\density{l} + \sqrt{\viscosity{v}}\density{v})},
\end{equation*}
where $\viscosity{l}, \viscosity{v}$ are the kinematic viscosities of
the liquid and vapor phase \cite{Miller1968}.
\begin{figure}[!ht]
  \centering
  \includegraphics[width=0.45\textwidth]{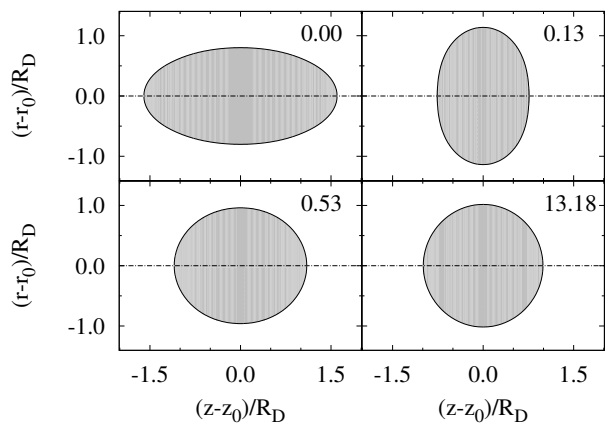}
  \caption{Time evolution of the shape of an ellipsoidal droplet
    immersed in a fluid with different density. Simulation parameters:
    $\systemsize = 320 \times 128$, $\interactionparameter =
    -6$. $\density{l}$ = 2.65 , $\density{v}$ = 0.075, $\dropcenter =
    (160.0, 0.5)$. Labels indicate the time corresponding to the
    different droplet shapes.}
  \label{fig:R9-ellipsoid-time-evolution}
\end{figure}

\begin{figure}[!ht]
  \centering
  \includegraphics[width=0.45\textwidth]{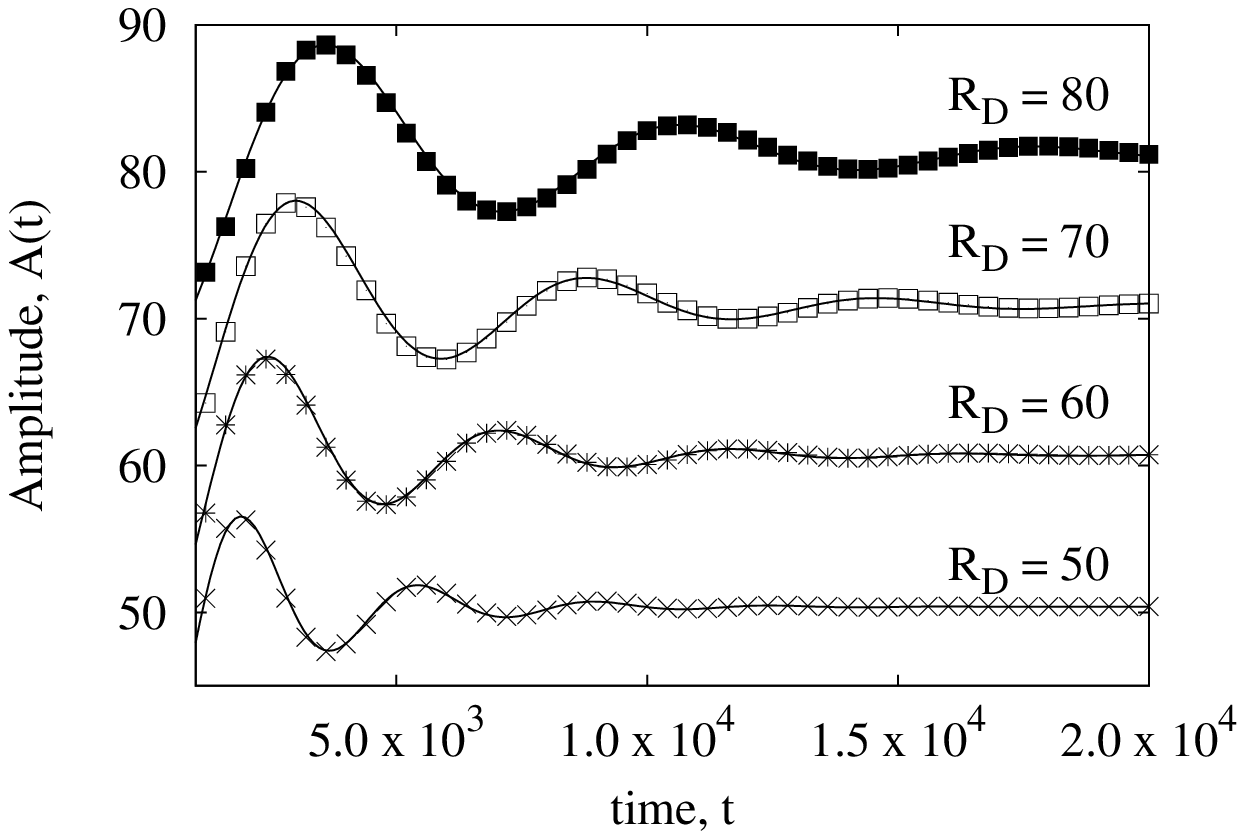}\\
  \includegraphics[width=0.45\textwidth]{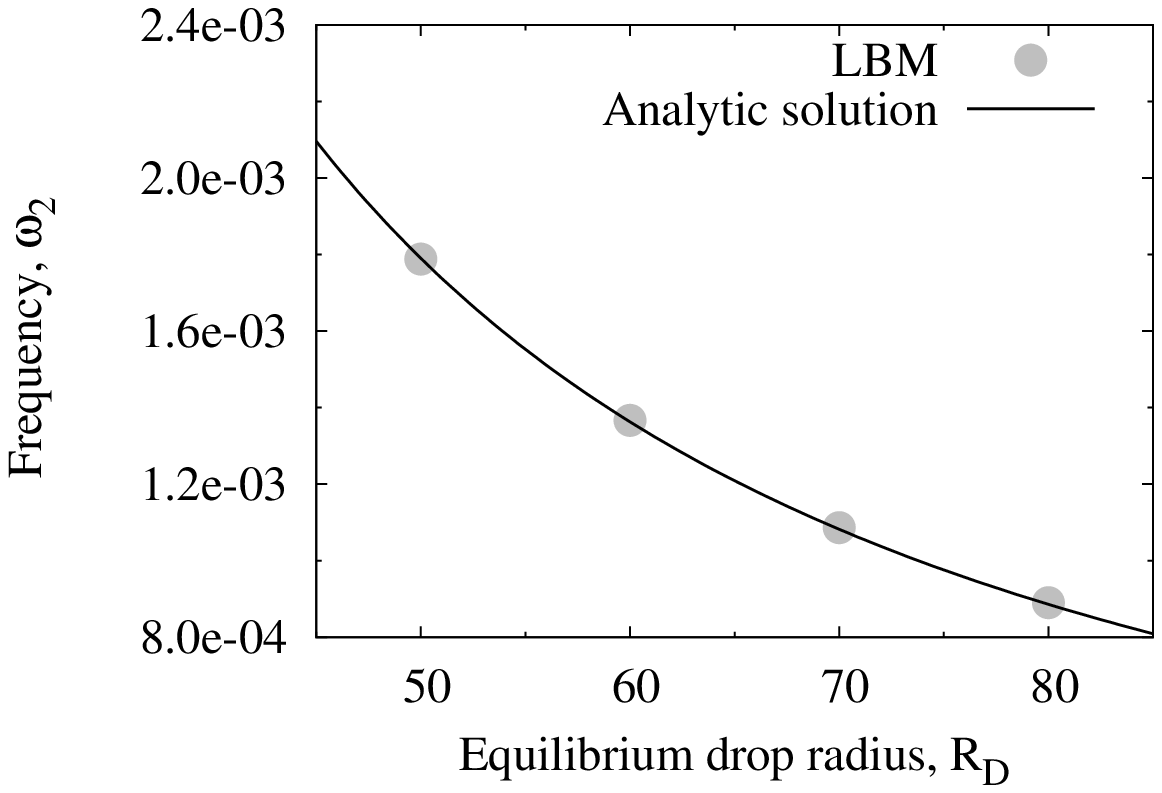}
  \caption{Top panel: amplitude, $A(t)$, of the oscillations 
    versus time, $t$, for different equilibrium droplet radii,
    $\dropradius$. Solid lines are obtained by fitting the function, ${g(t) = \dropradius  +  a\exp(-b t)\sin(\omega_2 t + d),}$
    to the data obtained from LBM
    simulations. Bottom panel: dimensionless frequency of the second
    mode of oscillation, $\omega_2$, versus the equilibrium droplet
    radius, $\dropradius$. Simulation parameters: $\systemsize= 320 \times
    128$, $\interactionparameter = -6$.}
  \label{fig:R9-oscillating-drop-test2}
\end{figure}  

In the LBM simulations for this test we use the free-slip boundary condition
at the top boundary and periodic boundary conditions 
at the left and right boundaries. The LBM simulations are
initialized with an axisymmetric ellipsoid, ${(z/R_a)^2 + (r/R_b)^2 = 1}$, where
$R_a, R_b$ are the intercepts on the $z$ and $r$-axis, respectively, 
with total volume $4\pi R_aR_b^2/3$. Due to the surface tension, the
ellipsoidal droplet oscillates and due to viscous damping it does
finally attain an equilibrium spherical shape with radius $\dropradius =
(R_aR_b^2)^{1/3}$ (due to volume conservation).  The time evolution of
one of these LBM simulations is shown in
FIG.~\ref{fig:R9-ellipsoid-time-evolution}.  The time is measured in the capillary time scale, ${t_{cap} = \sqrt{\dropradius^3\rho_l/\st}}$. The LBM simulations are
performed to validate the effect of the droplet size, $\dropradius$,
on the frequency of oscillation, $\omega_2$. In order to calculate the
frequency of the oscillation we first measure the length of the
intercept on the $r$-axis as a function of time, $A(t)$ with $A(t = 0) =
R_b$, and then we fit the function 
${g(t) = \dropradius  +  a\exp(-b t)\sin(\omega_2 t + d),}$
(see FIG.~\ref{fig:R9-oscillating-drop-test2}). We find that the
numerical estimation of the frequency of the oscillation of the
droplet is in excellent agreement with the theoretically expected
value, with a maximum relative error of approximatly 1\% (see FIG.~\ref{fig:R9-oscillating-drop-test2}).

\subsubsection{Rayleigh - Plateau (RP) instability}
The last problem that we consider for the validation is the breakup of
a liquid thread into multiple droplets. The problem was first
studied experimentally by Plateau \cite{Plateau1873} and later
theoretically by Lord Rayleigh \cite{Rayleigh1879}, and is currently
referred to as Rayleigh-Plateau (RP) instability. The RP instability
has been extensively studied experimentally, theoretically and
numerically \cite{Plateau1873,Rayleigh1879,Lafrance1975,Tomotika1935}. Moreover,
the problem is fully axisymmetric and therefore suitable for the
validation of our multiphase axisymmetric LBM model.

In this validation we check the instability criterion: a liquid
cylinder of radius $\cylinderradius$ is unstable, if the wavelength of
a disturbance, $\lambda$, on the surface of a liquid cylinder is longer then its circumference
${2\pi\cylinderradius}$. Moreover, we compare the radius of the
resulting drops with experimental \cite{Rutland1971} and numerical
data \cite{Driessen2011}.

For the LBM simulations we use free-slip boundary condition at the top
boundary and periodic boundary conditions at left and right
boundaries. The LBM simulations are performed in a domain of size ${\systemsize = \lambda \times 450}$.
The wavelength, $\lambda$, of the noise runs over 576, 768, 1024, 1280, 1536 and 1792 for different wavenumbers, $\kappa = 2\pi/\lambda$.
We represent the wavenumber in dimensionless form as $\kappa^* = \kappa \cylinderradius$.
The SC interaction parameter, ${G = -6.0}$, liquid density $\rho_l = 2.68$, vapor density $\rho_v = 0.078$, surface tension $\st = 0.141$ and kinematic viscosity $\nu = 0.016$ are 
fixed for these simulations. For these parameters the Ohnesorge number, ${Oh = \nu\sqrt{\rho_l/ (\st\cylinderradius}) = 0.09}$. The axial velocity field in the liquid cylinder is initialized by using sinusoidal velocity 
field as ${u_z(z, r) = \epsilon_u \sin(2\pi z /\lambda)}$. For our LBM simulation we use $\epsilon_u < 5\times 10^{-3}$. 

\begin{figure}[!ht]
  \centering
  \includegraphics[width=0.42\textwidth]{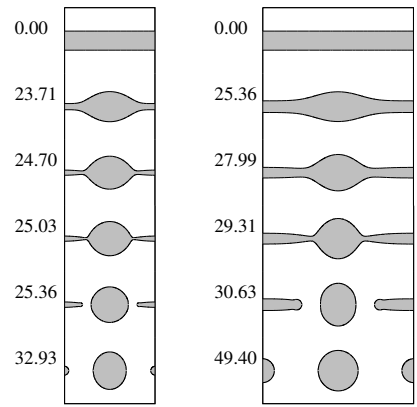}
  \caption{Growth of the Rayleigh-Plateau instability with time. Left
    panel: $\kappa^*$ = 0.65; right panel: $\kappa^*$ = 0.39. Labels on figures
    indicate the corresponding dimensionless time, $t/t_{cap}$.}
  \label{fig:rp-cylinder-breakup}
\end{figure}

\begin{figure}[!ht]
  \centering
  \includegraphics[width=0.46\textwidth]{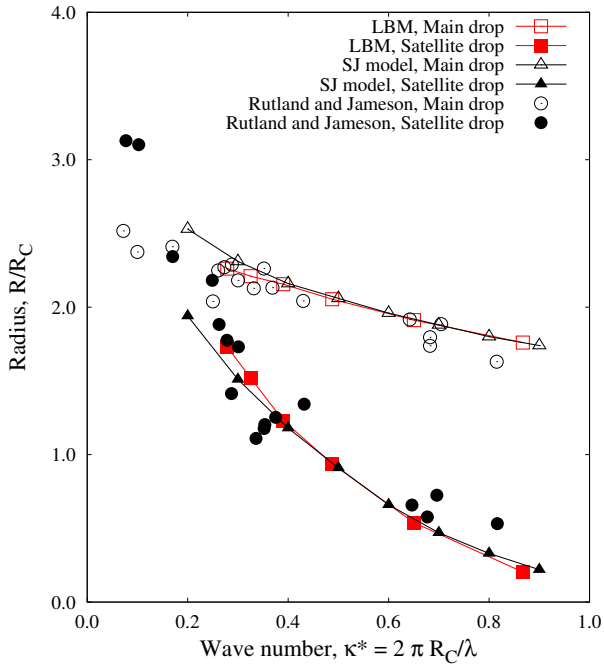}
  \caption{Dimensionless wave-number $\kappa^*$ versus dimensionless
    droplet radius, $ R/\cylinderradius$. Triangles markers
    represents the data obtained from the SJ model for $Oh = 0.1$ \cite{Driessen2011}, circle markers
    represents the data from experiments \cite{Rutland1971} and square markers represents 
    data from the axisymmetric LBM simulations for $Oh = 0.09$.}
  \label{fig:rp-phasedia}
\end{figure}

The time evolution of the RP instability corresponding to two
different wavenumber ${\kappa^* = 2\pi \cylinderradius/\lambda}$ is shown in
FIG.~\ref{fig:rp-cylinder-breakup}. The time is measured in the capillary time scale, ${t_{cap} = \sqrt{\cylinderradius^3\rho_l/\st}}$. In our simulations we find that
the cylinder breaks up into two or more droplets as long as the
condition $\kappa^* < 1$ is satisfied (corresponding to the RP instability
criterion, ${2\pi \cylinderradius < \lambda}$). Furthermore, the comparisons of drop
sizes for different wavenumber shown in FIG.~\ref{fig:rp-phasedia} is
in excellent agreement with the results of the slender jet approximation model (SJ)  \cite{Driessen2011} and with 
experimental data \cite{Rutland1971}.

\section{Conclusions}\label{sec:conclusion}
In the present manuscript we introduced a novel axisymmetric LBM
formulation that can be employed for single-phase as well as for
multiphase flows. The multiphase model is the widely employed
Shan-Chen model and the axisymmetric version here described is
particularly convenient as it allows one to easily switch from 3D to
2D axisymmetric simulations while maintaing the usual Shan-Chen
parameters (i.e.  densities and coupling strength).  The lattice
Boltzmann axisymmetric model allows for the solution of multiphase
flows at the computational cost of a 2D simulation. One particular
interesting application comes from the possibility of increasing the
system size, thus reducing the relative size of the LBM diffuse
interface with respect to all other length scales in the flow. We
presented several validations for single-phase as well as for
multiphase flows. In the case of multiphase flows we have
quantitatively validated the mass conservation and the dynamics of an
axially symmetric oscillating droplet. \REM{The present model can be used
to study the dynamics of system such as the breaking of a liquid
thread into main and secondary droplets, a problem of interested, for
example, for the formation of droplets from ink-jet printer nozzles.}
\ADD{We have also successfully validated the present model for the contraction of viscous ligament for $Re \sim 7$ and $We \sim 1$ \cite{Srivastava2013}. Currently, we are employing the model to study the flow from ink-jet nozzles at $Re \sim 100$ and $We \sim 50$. Of course there may be limitations on the use of any type of axisymmtric model concerning higher Re and We number flows. This concerns the case where velocity and/or density may develop fluctuation that break the axisymmetry. These type of flows clearly can never be captured by axisymmetric flows solver.}
The constraint of axis-symmetry may partially be relaxed by models
that keep into account azimuthal perturbation to lowest order, this
will be the subject of future work.

\section*{Acknowledgments}
We acknowledge useful discussion with Roger Jeurissen, Theo
Driessen and Luca Biferale.  This work is part of the research program of the Foundation
for Fundamental Research on Matter (FOM), which is part of the
Netherlands Organization for Scientific Research (NWO).

\bibliographystyle{prsty}
\bibliography{reference}

\clearpage
\newpage

\appendix
\section{Chapman-Enskog on modified LBM}\label{sec:appA}
The modified lattice Boltzmann Eq.~(\ref{eq:lbm2d_mod_new}) for the distribution function $f_i(\vect{x},t)$ reads
\begin{equation}\label{eq:lat_bolt_eq}
  \begin{split}
    f_i({\vect{x}}+ {\vect{c}_i}\delta t, t + \delta t) &- f_i({\vect{x}}, t) = -\frac{1}{\tau} \big(f_i({\vect{x}},t)-\feq_i(\rho, \ueq)\big) \\
    &~~~~+ \delta t ~h_i(\vect{x} + {\vect{c}_i} \delta t /2, t + \delta t/2).
  \end{split}
\end{equation}
where $h_i$ is the source terms, $\vect{c}_i$ is the lattice velocities $\tau$ is the relaxation parameter and $\feq_i$ is the discrete second order approximation of the Maxwell-Boltzmann distribution function 
\begin{eqnarray}\label{eq:e_distribution}
  \feq_i(\rho,\vect{u}) &=& W_i\rho\bigg[ 1 + \frac{1}{c_s^2}(\vect{c}_i\cdot\vect{u}) + \frac{1}{2c_s^4} (\vect{c}_i\cdot\ueq)^2 \nonumber\\
    && - \frac{1}{2c_s^2}|\ueq|^2\bigg],
\end{eqnarray}
where $c_s$ is the speed of sound and $W_i$'s are the weight factors to ensure the symmetry of the lattice. For the D2Q9 LB model with BGK collision operator the speed of sound, $c_s = \sqrt{1/3}$, $W_0 = 4/9$, $W_i = 1/9$ for $i = 1,2,3,4$ and $W_i = 1/36$ for $i = 5,6,7,8$. 
In general these weights \ADD{are positive and} satisfy following symmetry \REM{relations} \ADD{conditions \cite{Wolf-gladrow2005}}
\newenvironment{equations}{\equation\aligned}{\endaligned\endequation}

\begin{equations}\label{eq:weights}
  \sum_i{W_i}       			&= 1, 					&	 \sum_i{W_i\cia}     &=0, \\
  \sum_i{W_i\cia\cib}			&= c_s^2\delta_{\alpha\beta},  	 &\sum_i{W_i\cia\cib\cig} &= 0, \nonumber
\end{equations}
\vskip -0.5cm
\begin{equations}
  &\sum_i{W_i\cia\cib\cig\cid} 	= c_s^4\big(\delta_{\alpha\beta}\delta_{\gamma\delta} + \delta_{\alpha\gamma}\delta_{\beta\delta} + \delta_{\alpha\delta}\delta_{\beta\gamma}),\\
  &\sum_i{W_i\cia\cib\cig\cid\cie} =0.
\end{equations}
The \REM{macroscopic} \ADD{hydrodynamic} density, $\rho$ and momentum, ($\rho \vect{u}$) are given by the
zeroth and first moment of the distribution function respectively, $i.e.,$

\begin{subequations}\label{eq:hydro_var}
  \begin{align}
    \rho(\vect{x},t) &= \sum_i{f_i(\vect{x},t)}, \\
    (\rho\vect{u})(\vect{x},t) &= \sum_i{\vect{c}_if_i(\vect{x},t)}.
  \end{align}
\end{subequations}
In absence of any external force, ${\ueq = \vect{u}}$.
In order to establish a relation between the LB Eq.~(\ref{eq:lat_bolt_eq}) continuity Eq.~(\ref{eq:axis_cty}) and the NS equations (\ref{eq:axis_ns}) it is necessary 
to separate different time scales. 
We distinguish between slow and fast varying quantities by using two time scales and one space scale \cite{Wolf-gladrow2005}. We expand the time and space derivative ($\gradc:$ the gradient operator in the Cartesian coordinate system) using a \ADD{formal} parameter $\varepsilon$ \REM{(Knudsen number)} as 
\begin{align}\label{eq:exp_tx}
  \dt = \varepsilon \dta + \varepsilon^2 \dtb + \Order (\varepsilon^3),
  \gradc = \varepsilon \gradc^{(1)}  + \Order (\varepsilon^2),
\end{align}
and the distribution function, $f_i$ as 
\begin{equation}\label{eq:exp_fi}
  f_i = \fiz + \varepsilon \fia + \varepsilon^2 \fib + \Order(\varepsilon^3).
\end{equation}
\ADD{The expansion parameter is formal in the sense that it allows us to keep the track of the terms with different order of magnitude with respect to $\fiz$.}
The zeroth order contribution $\fiz$ is exactly the same as the equilibrium distribution function,$\feq_i$. The first and second order perturbations \ADD{in $f_i$} do not contribute to \REM{mass} \ADD{hydrodynamic density} and momentum \cite{Wolf-gladrow2005} :
\begin{subequations}\label{eq:per_assumprtion}
  \begin{align}
    \sum_i{\fia} &= \sum_i{\fib} = 0,\\
    \sum_i{\vect{c}_i\fia} &= \sum_i{\vect{c}_i\fib} = 0.
  \end{align}
\end{subequations}
The source term $h_i$ does not have any zeroth order contribution and is expanded as 
\begin{equation}
  h_i = \varepsilon \hia + \varepsilon^2 \hib + \Order(\varepsilon^3).
\end{equation}
\begin{widetext}
  Taylor series of $f_i$ and $h_i$ around $(\vect{x},t)$ are given by
  \begin{eqnarray}
    f_i(\vect{x} + \vect{c}_i\delta t,t+\delta t) &=& f_i(\vect{x},t) + \delta t (\dt + \cia \partial_\alpha)f_i(\vect{x},t) + \frac{(\delta t)^2}{2} (\dt + \cia \partial_\alpha)^2f_i(\vect{x},t) + \Order(({\delta t})^3), \label{eq:taylor_series_f} \\
    h_i(\vect{x} + \vect{c}_i\delta t/2, t + \delta t/2) &=& h_i(\vect{x},t) + \frac{\delta t}{2} (\dt + \cia \partial_\alpha)h_i(\vect{x},t) 
    +\frac{1}{2} \left(\frac{\delta t}{2}\right)^2 (\dt + \cia \partial_\alpha)^2h_i(\vect{x},t) + \Order(({\delta t})^3),\label{eq:taylor_series_h}
  \end{eqnarray}
  where $\cia$ is the $\alpha$-{\it th} component of $\vect{c}_i$, and $\partial_\alpha$ represents the partial derivative with respect to $\alpha$-{\it th} component of \vect{x}. Indices $\alpha, \beta, \gamma, \delta$ used in the following derivation ranges over the set $\{z,r\}$, and when an index appears twice in a single term it represents the standard Einstein summation convention. 
  Using Eq.~(\ref{eq:exp_tx}),(\ref{eq:exp_fi}),(\ref{eq:taylor_series_f}), and (\ref{eq:taylor_series_h}) in (\ref{eq:lat_bolt_eq}) and rearranging the terms we obtain a
  series in $\varepsilon$ 
  \begin{equation}\label{eq:main_collect}
    \begin{split}
      \varepsilon \left[ \delta t \left(\dta \fiz+ \cia \partial_\alpha^{(1)}\fiz\right)   \right] 
      +\varepsilon^2 \left[ \delta t \left(\dtb \fiz + \dta \fia + \cia \partial_\alpha^{(1)}\fia\right) +\frac{(\delta t)^2}{2}\left( \dta\dta \fiz 
        + \cia \cib \partial_\alpha^{(1)} \partial_\beta^{(1)}\fiz \right.\right.\\ \left.\left.
        + 2 \cia \dta \partial_\alpha^{(1)}\fiz \right)\right] 
      = \varepsilon \left[ -\frac{1}{\tau} \fia + \delta t ~\hia \right]  + \varepsilon^2 \left[-\frac{1}{\tau} \fib + \delta t ~\hib + \frac{(\delta t)^2}{2}\Big(\dta + \cia\partial_\alpha^{(1)}\Big)\hia \right] + \Order(\varepsilon^3).
    \end{split}
  \end{equation}
  Comparing the coefficients of $\varepsilon$, $\varepsilon^2$ and omitting $\varepsilon^3$ terms in Eq.~(\ref{eq:main_collect}) gives us 
  \begin{equation}\label{eq:coff_eps}
    \delta t \left(\dta \fiz+ \cia \partial_\alpha^{(1)}\fiz\right) =  -\frac{1}{\tau} \fia + \delta t ~\hia,
  \end{equation}
  \begin{equation}\label{eq:coff_eps2}
    \begin{split}
      \delta t \left(\dtb \fiz + \dta \fia + \cia \partial_\alpha^{(1)}\fia\right) +\frac{(\delta t)^2}{2}\left( \dta\dta \fiz + \cia \cib \partial_\alpha^{(1)} \partial_\beta^{(1)}\fiz + 2 \cia \dta \partial_\alpha^{(1)}\fiz \right) \\
      = -\frac{1}{\tau} \fib + \delta t ~\hib + \frac{(\delta t)^2}{2}\Big(\dta + \cia\partial_\alpha^{(1)}\Big)\hia,~~~~~~~~~
    \end{split}
  \end{equation}
  respectively. In the following steps of the CE expansion we will take the zeroth and first lattice velocity moments of Eqs.~(\ref{eq:coff_eps}) and (\ref{eq:coff_eps2}). The zeroth moment of Eqs.~(\ref{eq:coff_eps}) and (\ref{eq:coff_eps2}) will give us the mass conservation up to $\varepsilon$ and $\varepsilon^2$ order terms, respectively, and the first moment of Eqs.~(\ref{eq:coff_eps}) and (\ref{eq:coff_eps2}) will give us the momentum conservation up to $\varepsilon$ and $\varepsilon^2$ order terms, respectively. Finally by using Eq.~(\ref{eq:exp_tx}) we will obtain equations that conserves the hydrodynamic mass and momentum up to $\Order(\varepsilon^2)$ perturbations in $f_i$.

  \subsubsection*{\ADD{Mass conservation}}
  \ADD{The zeroth order moment is obtained by taking summation of  Eq.~(\ref{eq:coff_eps}) over index $i$, and the first order moment is obtained by multiplying  Eq.~(\ref{eq:coff_eps}) by ${\bf c}_i$ and taking summation over index $i$.} The zeroth and first order moments of Eq.~(\ref{eq:coff_eps}) along with Eqs.~(\ref{eq:hydro_var}) and (\ref{eq:per_assumprtion}) \REM{give} \ADD{gives us }
  \begin{align}
    \dta \rho + \partial_\alpha^{(1)} (\rho u_\alpha) &= \sum_i \hia, \label{eq:zeromom_eps}\\
    \dta (\rho u_\beta) + \partial_\alpha^{(1)} \Pi_{\alpha\beta}^{(0)} &= \sum_i \cib\hia \label{eq:firstmom_eps0}, 
  \end{align}
  \ADD{respectively}. \REM{where} $\Pi_{\alpha\beta}^{(0)}$ \ADD{in Eq.~(\ref{eq:firstmom_eps0})} is the zeroth order stress tensor, and \REM{defined as} \ADD{using Eq.~(\ref{eq:e_distribution}) it can be expressed in term of hydrodynamic variables \cite{Wolf-gladrow2005}}
  \begin{equation}\label{eq:ST}
    \Pi_{\alpha\beta}^{(0)} \equiv \sum_i\cia\cib\fiz  = \rho\left(c_s^2\delta_{\alpha\beta}   + u_\alpha u_\beta \right). 
  \end{equation}
  \ADD{Using Eq.~(\ref{eq:ST}) in Eq.~(\ref{eq:firstmom_eps0}) gives us
    \begin{equation}
      \dta (\rho u_\beta) + \partial_\alpha^{(1)}(\rho u_\alpha u_\beta)  = - \db \big(c_s^2\rho\big) + \sum_i \cib\hia \label{eq:firstmom_eps}, 
    \end{equation}
  } 
  \ADD{Eq.~(\ref{eq:zeromom_eps}) gives us the density change in convective time scale. In order to estimate the density change in during the diffusive process, we take the zeroth moment of Eq.~(\ref{eq:coff_eps2})}
  \REM{Next, we calculate the zeroth and first order moments of Eq.~(\ref{eq:coff_eps2}). The zeroth moment of Eq.~(\ref{eq:coff_eps2}) gives}
  \begin{equation*}
    \begin{split}
      \delta t \bigg(\dtb \sum_i\fiz + \dta \sum_i\fia + \partial_\alpha^{(1)}\sum_i\cia\fia\bigg) +\frac{(\delta t)^2}{2}\bigg( \dta\dta \sum_i\fiz +  \partial_\alpha^{(1)} \partial_\beta^{(1)}\sum_i\cia \cib\fiz \\
      + 2 \dta \partial_\alpha^{(1)}\sum_i\cia\fiz \bigg) 
      = -\frac{1}{\tau} \sum_i\fib + \delta t \sum_i\hib + \frac{(\delta t)^2}{2}\bigg(\dta\sum_i\hia + \partial_\alpha^{(1)}\sum_i\cia\hia\bigg).~~~~~~~~~
    \end{split}
  \end{equation*}
  Using Eqs.~(\ref{eq:per_assumprtion}), (\ref{eq:hydro_var}) and (\ref{eq:ST}) we get
  \begin{eqnarray*}
    \dtb \rho  +\frac{\delta t}{2}\left( \dta\dta \rho +  \partial_\alpha^{(1)} \partial_\beta^{(1)}\Pi_{\alpha\beta}^{(0)} + 2 \dta \partial_\alpha^{(1)}\big(\rho u_\alpha\big) \right) &=&  \sum_i\hib + \frac{\delta t}{2}\Big(\dta\sum_i\hia + \partial_\alpha^{(1)}\sum_i\cia\hia\Big).\\
    \dtb \rho  +\frac{\delta t}{2}\bigg( \dta\left(\dta \rho +\partial_\alpha^{(1)}(\rho u_\alpha)\right) +  \partial_\alpha^{(1)} \left( \dta (\rho u_\alpha) + \partial_\beta^{(1)}\Pi_{\alpha\beta}^{(0)} \right)\bigg) &=&  \sum_i\hib + \frac{\delta t}{2}\bigg(\dta\sum_i\hia + \partial_\alpha^{(1)}\sum_i\cia\hia\bigg).\\
  \end{eqnarray*}
  Finally using Eq.~(\ref{eq:zeromom_eps}) and (\ref{eq:firstmom_eps}) we get 
  \begin{eqnarray} \label{eq:zeromom_eps2_0}
    \dtb \rho  +\frac{\delta t}{2}\bigg( \dta\sum_i\hia +  \partial_\alpha^{(1)} \sum_i \cia \hia\bigg) &=&  \sum_i\hib + \frac{\delta t}{2}\bigg(\dta\sum_i\hia + \partial_\alpha^{(1)}\sum_i\cia\hia\bigg).
  \end{eqnarray}
\end{widetext}
Rearranging the terms of Eq.~(\ref{eq:zeromom_eps2_0}) gives \ADD{us rate of change of density density with diffusive time scale}
\begin{eqnarray} \label{eq:zeromom_eps2}
  \dtb \rho &=& \sum_i \hib.
\end{eqnarray}
We assume that the source term $h_i^2$ does not change the density at diffusive time scale, $i.\,e.$
\begin{equation}\label{eq:corr_ns_eqn_0}
  \sum_i \hib = 0.
\end{equation}
Using the relation $\varepsilon$(\ref{eq:zeromom_eps})+ $\varepsilon^2$ (\ref{eq:zeromom_eps2}) we get
\begin{equation}\label{eq:cty_eps01}
  \dt \rho + \partial_\alpha (\rho u_\alpha) = \varepsilon \sum_i\hia + \varepsilon^2 \sum_i\hib.
\end{equation}
If we choose 
\begin{eqnarray}
  \varepsilon\hia &=& -\frac{W_i \rho u_r }{r},\label{eq:corr_cty_eqn}
\end{eqnarray}
then  
\begin{subequations}
  \begin{align}\label{eq:corr_cty_eqn2}
    \sum_i\hia &= -\frac{1}{\varepsilon}\frac{\rho u_r}{r},\\
    \sum_i\cia\hia &= 0, \\
    \sum_i\cia\cib\hia &= -c_s^2\frac{1}{\varepsilon}\frac{\rho u_r}{r} \delta_{\alpha\beta}.
  \end{align}
\end{subequations}
\REM{Furthermore, if we choose
  $$\sum_i \hib = 0,$$
  then} \ADD{Using} the Eqs.~(\ref{eq:cty_eps01}), (\ref{eq:corr_cty_eqn2}) and (\ref{eq:corr_ns_eqn_0}) gives us
\begin{equation}\label{eq:cty_eps}
  \dt \rho + \partial_\alpha (\rho u_\alpha) = -\frac{\rho u_r}{r}.
\end{equation}
Eq.~(\ref{eq:cty_eps}) is the axisymmetric continuity Eq.~(\ref{eq:axis_cty}).\\
%
\begin{widetext}

  \subsubsection*{\ADD{Momentum conservation}}
  \ADD{Similarly, in order to calculate the rate of momentum change with respect to diffusive time scale,}  \REM{Now} we take the first moment of Eq.~(\ref{eq:coff_eps2}) 
  \begin{equation*}
    \begin{split}
      \delta t \bigg(\dtb \sum_i\cig\fiz + \dta \sum_i\cig\fia + \partial_\alpha^{(1)}\sum_i\cia\cig\fia\bigg) +\frac{(\delta t)^2}{2}\bigg( \dta\dta \sum_i\cig\fiz +  \partial_\alpha^{(1)} \partial_\beta^{(1)}\sum_i\cia \cib\cig\fiz \\
      + 2 \dta \partial_\alpha^{(1)}\sum_i\cia\cig\fiz \bigg) 
      = -\frac{1}{\tau} \sum_i\cig\fib + \delta t~\sum_i \cig\hib + \frac{(\delta t)^2}{2}\bigg(\dta \sum_i \cig \hia + \frac{\delta t}{2} \partial_\alpha^{(1)} \sum_i\cig\cia\hia\bigg),~~~~~~~~~
    \end{split}
  \end{equation*}
  using Eqs.~(\ref{eq:hydro_var}), (\ref{eq:per_assumprtion}) and (\ref{eq:corr_cty_eqn2}) we get
  \begin{equation}
    \begin{split}
      \dtb (\rho u_\gamma) + \partial_\alpha^{(1)} \Pi_{\alpha\gamma}^{(1)} + \frac{\delta t}{2}\left( \dta \dta (\rho u_\gamma) + \partial_\alpha^{(1)} \partial_\beta^{(1)}P_{\alpha\beta\gamma}^{(0)} + 2 \dta \partial_\alpha^{(1)} \Pi_{\alpha\gamma}^{(0)}\right) 
      =\sum_i \cig\hib - c_s^2\frac{1}{\varepsilon}\partial_\gamma^{(1)}\Big(\frac{\rho u_r}{r} \Big), \label{eq:firstmom_eps2} 
    \end{split}
  \end{equation}
\end{widetext}
where 
\begin{align}	                
  P_{\alpha\beta\gamma}^{(0)}   & \equiv \sum_i\cia\cib\cig\fiz,\label{eq:pp_tensor}\\
  \Pi_{\alpha\gamma}^{(1)} & \equiv \sum_i{\cia\cig \fia}.\label{eq:ss_tensor}
\end{align}
Using Eq.~(\ref{eq:e_distribution}) and (\ref{eq:weights}) in Eq.~(\ref{eq:pp_tensor}) we get
\begin{align}\label{eq:Tensor_c}
  P_{\alpha\beta\gamma}^{(0)}  & = \frac{1}{c_s^2}\sum_i{W_i\cia\cib\cig\cid (\rho u_\delta)} \nonumber\\
  & = c_s^2 \Big(\delta_{\alpha\beta} (\rho u_\gamma)+ \delta_{\beta\gamma} (\rho u_\alpha)+ \delta_{\alpha\gamma} (\rho u_\beta)\Big),	
\end{align}
and Eq.~(\ref{eq:coff_eps}) in Eq.~(\ref{eq:ss_tensor}) gives
\begin{align}
  \Pi_{\alpha\gamma}^{(1)}& = \delta t~\tau\sum_i{\cia\cig\left(\hia - \cid\partial_\delta^{(1)}\fiz - \dta\fiz\right)} \label{eq:Tensor_b}\nonumber\\
  & = \delta t~\tau\sum_i \cia\cig \hia   - \delta t~\tau \left( \partial_\delta^{(1)} P^{(0)}_{\alpha\gamma\delta} + \dta \Pi^{(0)}_{\alpha\gamma}\right).
\end{align}
Substituting Eqs.~(\ref{eq:firstmom_eps}) and (\ref{eq:Tensor_b}) in (\ref{eq:firstmom_eps2}) and rearranging gives
\begin{equation}\label{eq:derivation3}
  \begin{split}
    &\dtb (\rho u_\gamma) - \delta t\left(\tau - \frac{1}{2}\right) \left( \partial_\alpha^{(1)} \partial_\delta^{(1)} P^{(0)}_{\alpha\gamma\delta} +\dta \partial_{\alpha}^{(1)} \Pi_{\alpha\gamma}^{(0)} \right) \\
    &= c_s^2\frac{1}{\varepsilon}\delta t \left(\tau - \frac{1}{2}\right)\partial_\gamma^{(1)}\Big(\frac{\rho u_r}{r} \Big) + \sum_i \cig\hib .
  \end{split}
\end{equation}

In order to obtain the NS Eq.~(\ref{eq:axis_ns}) from the lattice Boltzmann Eq.~(\ref{eq:lat_bolt_eq}) it is necessary that the
hydrodynamic velocity \REM{should satisfy} \ADD{satisfies} the low Mach number, $Ma$ condition $i.e.~\Order(Ma^3)$ terms are very small and can be neglected from the Eq.~(\ref{eq:derivation3}). \ADD{For LB method the Mach number is defined as $Ma = u/\cs$, where $u$ is the characteristic hydrodynamic velocity and $\cs$ is the speed of sound in LB method.} The third order velocity appears only in the expression  $\dta \partial_{\alpha}^{(1)} \Pi_{\alpha\gamma}^{(0)}$ in Eq.~(\ref{eq:derivation3}):
\begin{equation*}
  \begin{split}
    \dta \partial_{\alpha}^{(1)} \Pi_{\alpha\gamma}^{(0)} &=\dta\left(\daone \Pi^{(0)}_{\alpha\gamma}\right) \nonumber\\
    & = \dta\left(\daone \left(\rho u_\alpha u_\gamma + c_s^2 \rho \delta_{\alpha\gamma}\right)\right) \\
    & = \dta\daone\big(\rho u_\alpha u_\gamma\big) + c_s^2 \dgone \big(\dta \rho\big) \\
    & = \daone\Big(\dtone (\rho u_\alpha) u_\gamma + \dtone (\rho u_\gamma) u_\alpha\\
    &~~   -(\dtone \rho)u_\alpha u_\gamma\Big) + c_s^2 \dgone \big(\dta \rho\big).
  \end{split}
\end{equation*}
Using Eqs. (\ref{eq:zeromom_eps}), (\ref{eq:firstmom_eps}) and (\ref{eq:corr_cty_eqn2}) we get  
\begin{equation*}
  \begin{split}
    &\dta \partial_{\alpha}^{(1)} \Pi_{\alpha\gamma}^{(0)} \\
    & = -\daone\bigg(\big(\dta \rho\big)u_\alpha u_\gamma + u_\gamma \partial_\beta^{(1)} \Pi_{\alpha\beta}^{(0)}  + u_\alpha\partial_\beta^{(1)} \Pi_{\gamma\beta}^{(0)}  \bigg) \\
    &~~- c_s^2 \dgone\Big(\dbone \big(\rho u_\beta\big) + \frac{1}{\varepsilon}\frac{\rho u_r}{r}\Big)\\
    & = -\daone\bigg(\big(\dta \rho\big)u_\alpha u_\gamma + u_\gamma \partial_\beta^{(1)} \Big(\rho u_\alpha u_\beta + c_s^2\rho \delta_{\alpha\beta}\Big) \\
    &~~ +u_\alpha\partial_\beta^{(1)} \Big(\rho u_\gamma u_\beta + c_s^2\rho \delta_{\gamma\beta}\Big) \bigg) \\
    &~~- c_s^2 \dgone\Big(\dbone \big(\rho u_\beta\big) + \frac{1}{\varepsilon}\frac{\rho u_r}{r}\Big)\\
    & = -\daone\bigg(\big(\dta \rho\big)u_\alpha u_\gamma + u_\gamma\partial_\beta^{(1)} \big(\rho u_\alpha u_\beta\big) + u_\alpha\partial_\beta^{(1)} \big(\rho u_\gamma u_\beta\big) \\
    &~~ + c_s^2\Big(u_\gamma \big(\partial_\alpha^{(1)}\rho\big) + u_\alpha\big(\partial_\gamma^{(1)} \rho\Big) \bigg) \\
    &~~- c_s^2 \dgone\Big(\dbone \big(\rho u_\beta\big) + \frac{1}{\varepsilon}\frac{\rho u_r}{r}\Big).\\
  \end{split}
\end{equation*}
Neglecting the terms $u_\alpha \dbone (\rho u_\beta u_\gamma)$, $(\dtone \rho)u_\alpha u_\gamma$ and $u_\gamma \dbone (\rho u_\beta u_\alpha)$ ($\Order(Ma^3)$ terms) from the last equation we get
\begin{equation}\label{eq:term2}
  \begin{split}
    \dta \partial_{\alpha}^{(1)} \Pi_{\alpha\gamma}^{(0)} & = -c_s^2\daone\bigg(u_\gamma\big(\partial_\alpha^{(1)} \rho\big) + u_\alpha\big(\partial_\gamma^{(1)} \rho\big) \bigg)\\
    &~~ - c_s^2 \dgone\Big(\dbone \big(\rho u_\beta\big) + \frac{1}{\varepsilon}\frac{\rho u_r}{r}\Big).
  \end{split}
\end{equation}

Hence using Eqs. (\ref{eq:Tensor_c}) and (\ref{eq:term2}), the second term on L.H.S. of Eq.~(\ref{eq:derivation3}) becomes
\begin{equation}
  \begin{split}
    &\partial_\alpha^{(1)} \partial_\delta^{(1)} P^{(0)}_{\alpha\gamma\delta} +\dta \partial_{\alpha}^{(1)} \Pi_{\alpha\gamma}^{(0)} \nonumber\\
    & =c_s^2 \Big( \ddone\ddone \big(\rho u_\gamma \big) + 2\ddone\dgone \big(\rho u_\delta\big) \Big) - c_s^2\daone\Big(u_\gamma\big(\daone \rho\big) \\
    &+  u_\alpha\big(\dgone \rho\big) + \dgone \big(\rho u_\beta\big) \Big)  - c_s^2\frac{1}{\varepsilon}\dgone \Big(\frac{\rho u_r}{r}\Big),
  \end{split}
\end{equation}
rearranging the terms we get
\begin{equation}
  \begin{split}
    &\partial_\alpha^{(1)} \partial_\delta^{(1)} P^{(0)}_{\alpha\gamma\delta} +\dta \partial_{\alpha}^{(1)} \Pi_{\alpha\gamma}^{(0)}\\
    &= c_s^2 \Big( \ddone\ddone \big(\rho u_\gamma\big) + 2\ddone\dgone \big(\rho u_\delta\big) - \dbone \big(u_\gamma \daone \rho\big) \\
    &~~ - \dbone\big( u_\alpha \dgone \rho\big) - \dbone\dgone \big(\rho u_\beta\big) \Big) - c_s^2\frac{1}{\varepsilon}\dgone \Big(\frac{\rho u_r}{r}\Big) \\
    &= c_s^2 \Big( \ddone\ddone \big(\rho u_\gamma\big) + \ddone\dgone \big(\rho u_\delta\big) \\
    &~~- \dbone \big(u_\gamma \daone \rho\big) - \dbone\big( u_\alpha \dgone \rho\big) \Big) - c_s^2\frac{1}{\varepsilon}\dgone \Big(\frac{\rho u_r}{r}\Big)\\
    &=c_s^2 \left( \ddone \big(\rho \ddone u_\gamma\big) + \ddone \big(\rho \dgone u_\delta\big)\right) - c_s^2\frac{1}{\varepsilon}\dgone \Big(\frac{\rho u_r}{r}\Big).\label{eq:derivation4}\\
  \end{split}
\end{equation}
Substituting Eq.~(\ref{eq:derivation4}) back in to Eq.~(\ref{eq:derivation3}) gives us
\begin{equation}\label{eq:derivation5}
  \begin{split}
    &\dtb \big(\rho u_\gamma\big) - c_s^2 \delta t\left(\tau - \frac{1}{2}\right) \bigg(\ddone \big(\rho \ddone u_\gamma\big) + \ddone \big(\rho \dgone u_\delta\big)\bigg)\\
    &+c_s^2 \frac{1}{\varepsilon}\delta t\left(\tau - \frac{1}{2}\right)\dgone \Big(\frac{\rho u_r}{r} \Big) \\
    &= c_s^2\frac{1}{\varepsilon}\delta t \left(\tau - \frac{1}{2}\right)\partial_\gamma^{(1)}\Big(\frac{\rho u_r}{r} \Big)  + \sum_i \cig\hib.
  \end{split}
\end{equation}
Using Eqs.~(\ref{eq:corr_cty_eqn2}) and rearranging we get 
\begin{eqnarray}\label{eq:derivation5}
  \dtb \big(\rho u_\gamma\big) &=& c_s^2 \delta t\left(\tau - \frac{1}{2}\right) \ddone \bigg(\rho \Big( \ddone u_\gamma + \dgone u_\delta\Big)\bigg)  \nonumber\\
  &&+ \sum_i \cig\hib .
\end{eqnarray}
Using relation $\varepsilon$ (\ref{eq:firstmom_eps}) + $\varepsilon^2$ (\ref{eq:derivation5}) along with Eq.~(\ref{eq:exp_tx}) we get
\begin{equation}\label{eq:der_ns_eq1}
  \begin{split}
    &\dt (\rho u_\gamma) + \partial_\alpha(\rho u_\alpha u_\gamma )  = - \partial_\gamma \big(c_s^2 \rho\big) \\
    &+  c_s^2 \delta t\left(\tau - \frac{1}{2}\right) \partial_\delta \bigg(\rho \Big( \partial_\delta u_\gamma + \partial_\gamma u_\delta\Big)\bigg)  + \varepsilon^2 \sum_i \cig\hib.
  \end{split}
\end{equation}
If we define $\viscosity{} = c_s^2\delta t(\tau - 0.5)$ and $p = c_s^2\rho$ Eq.~(\ref{eq:der_ns_eq1}) becomes
\begin{eqnarray}\label{eq:2der_ns_eq}
  \dt (\rho u_\gamma) + \partial_\alpha(\rho u_\alpha u_\gamma ) &=&- \partial_\gamma p  + \viscosity{} \partial_\delta \bigg(\rho \Big( \partial_\delta u_\gamma + \partial_\gamma u_\delta\Big)\bigg) \nonumber\\
  &&   +\varepsilon^2\sum_i \cig\hib. 
\end{eqnarray}
Eq.~(\ref{eq:2der_ns_eq}) represents axisymmetric NS equation if the source term $\hib$ satisfies the following conditions :
\begin{eqnarray}
  \varepsilon^2\sum_i\cir\hib &=&2\mu\dr\Big(\frac{u_r}{r}\Big) 
  - \frac{\rho u_r^2}{r} ,\label{eq:corr_r_ns_eqn}\\
  \varepsilon^2\sum_i\ciz\hib &=& \frac{\mu}{r}(\dr u_z + \dz u_r) 
  - \frac{\rho u_r u_z}{r} .\label{eq:corr_z_ns_eqn}
\end{eqnarray}
Finally we summarize the conditions on $\hia$ and $\hib$ that gives us axisymmetric NS equation in long wavelength and small Mach number limit:
\begin{align*}
  \sum_i \hia &= -\frac{1}{\varepsilon}\frac{\rho u_r}{ r},\\
  \sum_i\cir\hia & = 0, & \sum_i\ciz\hia & = 0,
\end{align*}
and
\begin{align*}
  \sum_i\hib &= 0,\\
  \sum_i\cir\hib &= \frac{1}{\varepsilon^2}\bigg(2\mu\dr\Big(\frac{u_r}{r}\Big) - \frac{\rho u_r^2}{r}\bigg),\\
  \sum_i\ciz\hib &=\frac{1}{\varepsilon^2}\bigg( \frac{\mu}{r}(\dr u_z + \dz u_r) - \frac{\rho u_r u_z}{r}\bigg) ,
\end{align*}
hence 
\begin{eqnarray*}
  h_i &=& \varepsilon \hia +  \varepsilon^2 \hib,\\
  &=& W_i\Big(-\frac{\rho u_r}{r} + \frac{1}{c^2_s}\big(c_{iz}H_z + c_{ir}H_r\big)\Big),
\end{eqnarray*}
which is the same as Eq.~(\ref{eq:H_terms}).
This ends our Chapman Enskog expansion procedure to obtain axisymmetric NS from modified LB equation
We do not impose any additional condition on density of fluid, $\rho$. 

\end{document}